%% file: pearl-jeffrey-data.tex
\documentclass[final,twoside,11pt]{entics} 
\usepackage{enticsmacro}

\newif\ifignore 
\ignorefalse
\newcommand{\auxproof}[1]{
\ifignore\mbox{}\newline
\textbf{PROOF:} \dotfill\newline
{\it #1}\mbox{}\newline
\textbf{ENDPROOF}\dotfill
\fi}

\usepackage{graphicx}
\usepackage{etex} 
\usepackage{amsmath}
\usepackage{amssymb}
\usepackage{amstext}
\usepackage{nccmath}
\usepackage{mathtools}
\usepackage{mathrsfs}
\usepackage{stmaryrd}
\usepackage{xspace}
\usepackage{enumitem}
\usepackage{cmll}
\usepackage{hyperref}
\usepackage{url}
\usepackage{nicefrac}
\usepackage{wrapfig}
\usepackage{pdfrender}
\usepackage{todonotes}

\usepackage{xypic}
\usepackage[all,cmtip]{xy}
\xyoption{tips}

\usepackage{tikzit}
\usetikzlibrary{circuits.ee.IEC}
\usetikzlibrary{shapes,shapes.geometric,shapes.misc}
\input{causality.tikzstyles}

\usepackage{color}
\definecolor{deepblue}{rgb}{0,0,0.5}
\definecolor{deepred}{rgb}{0.6,0,0}
\definecolor{deepgreen}{rgb}{0,0.5,0}
\definecolor{darkgray}{rgb}{0.5,0.5,0.5}

\usepackage{listings}

\DeclareFixedFont{\ttb}{T1}{txtt}{bx}{n}{9} 
\DeclareFixedFont{\ttm}{T1}{txtt}{m}{n}{9}  

\lstdefinestyle{python_ppl}{
	language=Python,
	basicstyle=\ttm,
	otherkeywords={let, to},          
	keywordstyle=\ttb\color{deepblue},
	emph={condition, Infer, =:=, observe, sample, score, normalize},     
	emphstyle=\ttb\color{deepred},    
	stringstyle=\color{deepgreen}  
}

\lstdefinelanguage{CustomML}{
	keywords={match, with, rec, true, false, fun, return, let, in, if, then, else, type, val, module, sig, end, ref, struct},
	keywordstyle=\color{deepblue}\bfseries,
	ndkeywords={ref},
	ndkeywordstyle=\color{darkgray}\bfseries,
	identifierstyle=\color{black},
	sensitive=false,
	comment=[l]{//},
	morecomment=[s]{/*}{*/},
	commentstyle=\color{darkgray}\ttfamily,
	stringstyle=\color{red}\ttfamily,
	morestring=[b]',
	morestring=[b]"
}

\lstdefinestyle{ml_ppl}{
	language=CustomML,
	basicstyle=\ttm,       
	keywordstyle=\ttb\color{deepblue},
	emph={Gauss,N,Infer,condition, =:=, observe, sample, score, normal, gp_sample},     
	emphstyle=\ttb\color{deepred},    
	stringstyle=\color{deepgreen}
}

\lstdefinelanguage{JavaScript}{
	keywords={typeof, new, true, false, catch, function, return, null, catch, switch, var, if, in, while, do, else, case, break},
	keywordstyle=\color{blue}\bfseries,
	ndkeywords={class, export, boolean, throw, implements, import, this},
	ndkeywordstyle=\color{darkgray}\bfseries,
	identifierstyle=\color{black},
	sensitive=false,
	comment=[l]{//},
	morecomment=[s]{/*}{*/},
	commentstyle=\color{darkgray}\ttfamily,
	stringstyle=\color{red}\ttfamily,
	morestring=[b]',
	morestring=[b]"
}

\lstdefinestyle{webppl}{
	language=JavaScript,
	basicstyle=\ttm,
	otherkeywords={let},            
	keywordstyle=\ttb\color{deepblue},
	emph={flip, condition, Infer, factor, sample, normal, score, observe},
	emphstyle=\ttb\color{deepred},    
	stringstyle=\color{deepgreen}
}

\lstdefinestyle{prolog}{
	language=Prolog,
	basicstyle=\ttm,         
	keywordstyle=\ttb\color{deepblue},
	emphstyle=\ttb\color{deepred},    
	stringstyle=\color{deepgreen}, 
}

\lstdefinestyle{funprolog}{
	language=Prolog,
	basicstyle=\ttm,
	otherkeywords={let, in, return},            
	keywordstyle=\ttb\color{deepblue},
	emphstyle=\ttb\color{deepred},    
	stringstyle=\color{deepgreen}, 
}

\newcommand*{\mlstinline}[1]{\text{\lstinline|#1|}}

\ifdefined\mathsl\else
  \DeclareMathAlphabet{\mathsl}{\encodingdefault}{\rmdefault}{\mddefault}{\sldefault}
  \SetMathAlphabet{\mathsl}{bold}{\encodingdefault}{\rmdefault}{\bfdefault}{\sldefault}
\fi

\newenvironment{myproof}{\begin{trivlist} \item[\hskip \labelsep%
{\bf Proof.}]}{\end{trivlist}}

\newcommand{\QEDbox}{\square}
\newcommand{\QED}{\hspace*{\fill}$\QEDbox$}

\newcommand*{\fatten}[1][.4pt]{%
  \textpdfrender{
    TextRenderingMode=FillStroke,
    LineWidth={\dimexpr(#1)\relax},
  }%
}

\ifdefined\mathsl\else
  \DeclareMathAlphabet{\mathsl}{\encodingdefault}{\rmdefault}{\mddefault}{\sldefault}
  \SetMathAlphabet{\mathsl}{bold}{\encodingdefault}{\rmdefault}{\bfdefault}{\sldefault}
\fi

\makeatletter
\newcommand{\mathoverlap}[2]{\mathpalette\mathoverlap@{{#1}{#2}}}
\newcommand{\mathoverlap@}[2]{\mathoverlap@@{#1}#2}
\newcommand{\mathoverlap@@}[3]{\ooalign{$\m@th#1#2$\crcr\hidewidth$\m@th#1#3$\hidewidth}}
\makeatother

\newcommand{\klafter}{\mathbin{\mathoverlap{\circ}{\cdot}}}

\DeclareSymbolFont{T1op}{T1}{cmr}{m}{n}
\SetSymbolFont{T1op}{bold}{T1}{cmr}{bx}{n}
\DeclareMathSymbol{\mathguilsinglleft}{\mathopen}{T1op}{'016}
\DeclareMathSymbol{\mathguilsinglright}{\mathclose}{T1op}{'017}

\newcommand{\idmap}[1][]{\ensuremath{\mathrm{id}_{#1}}}
\newcommand{\after}{\mathrel{\circ}}
\newcommand{\pull}{\mathrel{\mathchoice%
   {\scalebox{-0.5}[1]{$\gg=$}}
   {\scalebox{-0.5}[1]{$\gg{\kern-1.5ex}=$}}
   {\scalebox{-0.5}[1]{${\kern.5ex}\scriptstyle\gg{\kern-0.2ex}={\kern.5ex}$}}
   {\scalebox{-0.5}[1]{$\scriptscriptstyle\gg=$}}}}
\newcommand{\push}{\mathrel{\mathchoice%
   {\scalebox{-0.5}[1]{$=\ll$}}
   {\scalebox{-0.5}[1]{$={\kern-1.5ex}\ll$}}
   {\scalebox{-0.5}[1]{${\kern.5ex}\scriptstyle={\kern-0.2ex}\ll{\kern.5ex}$}}
   {\scalebox{-0.5}[1]{$\scriptscriptstyle=\ll$}}}}
\newcommand{\flrn}{\ensuremath{\mathsl{flrn}}}

\newcommand{\multinomial}[1][]{\ensuremath{\mathsl{mn}[#1]}}

\newcommand{\no}[1]{#1^{\scriptscriptstyle \bot}} 

\DeclareMathOperator*{\argmax}{argmax} 
\DeclareMathOperator*{\argmin}{argmin} 
\newcommand{\DKL}{\ensuremath{\mathsl{D}_{\mathsl{KL}}}}

\newcommand{\setin}[3]{\{#1\in#2\;|\;#3\}}
\newcommand{\supp}{\mathsl{supp}}
\newcommand{\acc}{\mathsl{acc}}
\newcommand{\arr}{\mathsl{arr}}

\newcommand{\Div}{\ensuremath{\mathfrak{D}{\kern-1.0pt}V}}
\newcommand{\coefm}[1]{\ensuremath{\fatten[0.6pt]{(}{\kern1pt}#1{\kern1pt}\fatten[0.6pt]{)}}}

\newcommand{\setsize}[1]{|{\kern.1em}#1{\kern.1em}|}
\newcommand{\bigsetsize}[1]{\big|{\kern.1em}#1{\kern.1em}\big|}

\newcommand{\ket}[1]{\ensuremath{|{\kern.1em}#1{\kern.1em}\rangle}}
\newcommand{\bigket}[1]{\ensuremath{\big|{\kern.1em}#1{\kern.1em}\big\rangle}}
\newcommand{\ketstrut}{\vrule height 10pt depth 5pt width 0pt}
\newcommand{\Bigket}[1]{\ensuremath{\left|\ketstrut{\kern.1em}#1{\kern.005em}\right>}}

\newcommand{\andthen}{\mathrel{\&}}
\newcommand{\bigandthen}{\mathop{\textnormal{\large\&}}}

\newcommand{\incr}[2]{#1\!\mathrel{\mathrm{+{\kern-.1em}+}}\!#2}

\newcommand{\distributionsymbol}{\mathcal{D}}

\newcommand{\multisetsymbol}{\mathcal{M}}

\newcommand{\Dst}{\distributionsymbol}

\newcommand{\Mlt}{\multisetsymbol}

\newcommand{\UF}{\ensuremath{\mathcal{U}{\kern-.75ex}\mathcal{F}}}

\newcommand{\Cat}[1]{\ensuremath{\mathbf{#1}}\xspace}

\newcommand{\Kl}{\mathcal{K}{\kern-.4ex}\ell}

\newcommand{\EM}{\mathcal{E}{\kern-.4ex}\mathcal{M}}

\newcommand{\Sets}{\Cat{Sets}}

\newcommand{\NNO}{\mathbb{N}}

\newcommand{\R}{\mathbb{R}}

\newcommand{\pR}{\R_{> 0}}

\newcommand{\Ef}{\ensuremath{\mathcal{E}{\kern-.5ex}f}}
\newcommand{\intd}{{\kern.2em}\mathrm{d}{\kern.03em}}
\newcommand{\indic}[1]{\mathbf{1}_{#1}}

\newcommand{\OF}{\ensuremath{\mathcal{O}{\kern-.1em}\mathcal{F}}}
\newcommand{\Closed}{\ensuremath{\mathcal{C}{\kern-.05em}\ell}}

\newsavebox\sbpto
\savebox\sbpto{\begin{tikzpicture}[baseline=-2.6pt]
            \filldraw[fill=white,draw=white] circle (1.0pt);
            \filldraw[fill=white,draw=black,line width=0.2pt] circle (1.2pt);
                \end{tikzpicture}}
\newcommand\chanto{\mathrel{\ooalign{$\rightarrow$\cr
            \hfil\!$\usebox\sbpto$\hfil\cr}}}

\newsavebox\sbground
\savebox\sbground{\begin{tikzpicture}[circuit ee IEC,yscale=1,xscale=1]
                \draw (0,-2ex) to (0,0ex) node[ground,rotate=90,xshift=.65ex] {};
                \end{tikzpicture}}

\newsavebox\sbunif
\savebox\sbunif{\begin{tikzpicture}[circuit ee IEC,yscale=1,xscale=1]
                \draw (0,0) to (0,2ex) node[ground,rotate=270,xshift=2.5ex] {};
                \end{tikzpicture}}

\tikzset{dot/.style =
  {inner sep=0mm,minimum width=1mm,minimum height=1mm,
    draw,shape=circle}}
\tikzset{minicopy/.style = {dot,fill,text depth=-0.2mm}}
\newsavebox\sbcopier
\savebox\sbcopier{%
  \begin{tikzpicture}[baseline=0pt]
    \node[minicopy,scale=.7] (a) at (0,3.6pt) {};
    \draw (a) -- +(-90:.30);
    \draw (a) -- +(45:.35);
    \draw (a) -- +(135:.35);
  \end{tikzpicture}}

\makeatother
 \newdir{ >}{{}*!/-7.5pt/@{>}}
 \newdir{|>}{!/4.5pt/@{|}*:(1,-.2)@^{>}*:(1,+.2)@_{>}}
 \newdir{ |>}{{}*!/-3pt/@{|}*!/-7.5pt/:(1,-.2)@^{>}*!/-7.5pt/:(1,+.2)@_{>}}

\def\conf{MFPS 2023} 	
\def\myconf{mfps39}
\volume{3}			
\def\volu{3}			
\def\papno{9}				

\def\lastname{Jacobs and Stein}
\def\runtitle{Pearl's and Jeffrey's Update as Modes of Learning in Probabilistic Programming} 
\def\mydoi{12281}
\def\copynames{B. Jacobs, D. Stein}    

\begin{document}

\lstset{style=webppl}

\begin{frontmatter}

\title{Pearl's and Jeffrey's Update as \\
Modes of Learning in Probabilistic Programming}

\author{Bart Jacobs \& Dario Stein}

\address{Institute for Computing and Information Sciences (iCIS) 
\\ 
Radboud University
\\
Nijmegen, The Netherlands
\\							
\href{mailto:bart@cs.ru.nl}{bart@cs.ru.nl}
\qquad
\href{mailto:dario.stein@ru.nl}{dario.stein@ru.nl}}


\begin{abstract} 
The concept of updating a probability distribution in the light of new
evidence lies at the heart of statistics and machine learning. Pearl's
and Jeffrey's rule are two natural update mechanisms which lead to
different outcomes, yet the similarities and differences remain
mysterious. This paper clarifies their relationship in several ways:
via separate descriptions of the two update mechanisms in terms of
probabilistic programs and sampling semantics, and via different
notions of likelihood (for Pearl and for Jeffrey). Moreover, it is
shown that Jeffrey's update rule arises via variational inference. In
terms of categorical probability theory, this amounts to an analysis
of the situation in terms of the behaviour of the multiset functor,
extended to the Kleisli category of the distribution monad.

\end{abstract}


\end{frontmatter}

\section{Introduction}\label{IntroSec}

Suppose you test for a certain disease, say Covid. You take three
consecutive tests, because you wish to be sure -- two of them come out positive but one is negative. How do you compute
the subsequent (posterior) probability that you actually have the
disease?  In a medical setting one starts from a prevalence, that is,
an \emph{a priori} disease probability, which is assumed to hold for
the whole population. Medical tests are typically not perfect: one has
to take their sensitivity and specificity into a account. They tell,
respectively, if someone has the disease, the probability that the test
is positive, and if someone does not have the disease, the probability
that the test is negative.

When all these probabilities (prevalence, sensitivity, specificity)
are known, one can apply Bayes' rule and obtain the posterior
probability after a single test. But what if we do three tests? And
what if we do a thousand tests?

It turns out that things become fuzzy when tests are repeated multiple
times. One can distinguish two approaches, associated with Pearl and
Jeffrey. They agree on single tests. But they may disagree wildly on
multiple tests, see the example in Section~\ref{ExSec} below. This is
disconcerting, certainly in the current age of machine learning, in
which so many decisions are based on statistical learning and decision
making.

Earlier work (of one of the authors)~\cite{Jacobs19c,Jacobs21c}
analysed the approaches of Pearl and Jeffrey. The difference there was
formulated in terms of learning from `what is right' and from `what is
wrong'. As will be recalled below, Pearl's update rule involves
increasing validity (expected value), whereas Jeffrey's rule involves
decreasing (Kullback-Leibler) divergence. The contributions of this
paper are threefold.
\begin{itemize}
\item It adds the perspective of probabilistic programming. Pearl's
  and Jeffrey's approaches to updating are formulated, for the medical
  test example, in a standard probabilistic programming language,
  namely WebPPL~\cite{dippl,probmods2}, see
  Section~\ref{ExSec}. Pearl's update is straightforwardly expressible
  using built-in conditioning constructs, while Jeffrey's update
  involves nested inference, a simple form of reasoning about
  reasoning~\cite{nestedinference}. We further explore the different
  dynamics behind the two update techniques are operationally using
  rejection samplers in Section~\ref{BouncerSec}.

\item The paper also offers a new perspective on the Pearl/Jeffrey
  distinction in terms of different underlying generative models and their associated likelihoods: with Pearl's
  update rule one increases one form of `Pearl' likelihood, whereas
  with Jeffrey's update rule one increases another form of `Jeffrey'
  likelihood. These two likelihoods are described in terms of
  different forms of evaluating data (as a multiset of data points)
  with respect to a multinomial distribution. Theses two forms of
  likelihood are directly related to the respective update mechanisms,
  see Section~\ref{LikelihoodSec}. Pearl likelihood occurs in
  practice, for example as the basis of the multinomial naive Bayes classifer~\cite{naivebayes}, while Jeffrey
  likelihood --- and its difference to Pearl's --- is new, as far as
  we know.

\item Pearl's likelihood directly leads to the associated update rule,
  see Theorem~\ref{PearlLikelihoodThm}. For Jeffrey's likelihood the
  connection is more subtle and involves variational
  inference~\cite{mackay2003information,Murphy12}: it is shown that
  Jeffrey's update is least divergent from the update rule for Jeffrey
  likelihood, in a suitable sense, see
  Theorem~\ref{JeffreyArgminThm}. This likelihood update rule is
  described categorically in terms of the extension of the multiset
  functor to the Kleisli category of the (discrete) distribution
  monad, see~\cite{DashS20,Jacobs21b}. This analysis clarifies the
  mathematical situation, for instance in
  Equation~\ref{ExtMltDagCommuteEqn}, where it is shown that this
  extended multiset functor commutes with the `dagger' reversal of
  channels. This is a new result, with a certain esthetic value.
\end{itemize}

\noindent This paper develops the idea that Pearl's and Jeffrey's rule
involve a difference in perspective: are we trying to learn something
about an individual or about a population?

\section{A Motivating Example}\label{ExSec}

Consider some disease with an \emph{a priori} probability (or
`prevalence') of $5\%$. There is a test for the disease with the
following characteristics:
\begin{itemize}
\item (`sensitivity') If someone has the disease, then the test is
  positive with probability of $90\%$.

\item (`specificity') If someone does not have the disease, there is a
  $95\%$ chance that the test is negative.
\end{itemize}

\noindent We are told that someone takes three consecutive tests
and sees two positive and one negative outcome. These test outcomes
are our observed data that we wish to learn from.

The question is: what is the posterior probability that this person
has the disease, in the light of this test data? You may wish to stop
reading here and calculate this probability yourself. Outcomes, using
Pearl's and Jeffrey's rule, will be provided in
Examples~\ref{PearlDiseaseEx} and~\ref{JeffreyDiseaseEx} below.

Below we present several possible implementations of the medical test
situation in the probabilistic programming language
WebPPL~\cite{dippl,probmods2}, giving three different solutions to the
above question. The code starts by defining a function \texttt{test}
which models the test outcome, incorporating the above sensitivity and
specificity. Here, \lstinline|flip(p)| tosses a biased coin with bias
\lstinline|p|.

\begin{lstlisting}
var test = function(dis) {
  return dis ? (flip(0.9) ? 'pos' : 'neg') : (flip(0.95) ? 'neg' : 'pos');
} 
\end{lstlisting}

\noindent We then define three inference functions which we simply
label as \texttt{prog1}, \texttt{prog2}, \texttt{prog3}. At this stage
we do not wish to connect them to Pearl/Jeffrey. We invite the reader
to form a judgement about what is the `right' way to model the above
situation with three test outcomes (`pos', `pos', `neg').

\begin{lstlisting}
var prog1 = function() {
  var dis = flip(0.05);  
  condition(test(dis) == 'pos');
  condition(test(dis) == 'pos');
  condition(test(dis) == 'neg');
  return dis; 
}

var prog2 = function() {
  var target = uniformDraw(['pos','pos','neg']);
  var dis = flip(0.05);
  condition(test(dis) == target);
  return dis;
}

var prog3 = function() {
  var target = uniformDraw(['pos','pos','neg']);
  return sample(Infer(function() {
    var dis = flip(0.05);
    condition(test(dis) == target);
    return dis;
  }))
}
\end{lstlisting}

\noindent All functions make use of the \lstinline|condition| command
to instruct WebPPL to compute a conditional
probability distribution. \texttt{prog1} uses three successive conditions, while
the other two use a single condition on a randomly chosen
\lstinline|target|. \texttt{prog3} additionally makes use of
\emph{nested inference}, that is, it wraps the \lstinline|Infer|
function around part of its code. Nested inference is a form of
reasoning about reasoning \cite{nestedinference} and has been applied
for example to the study of social cognition, linguistics and theory
of mind \cite[Ch.~6]{probmods2}. We give a short overview of WebPPL's semantics and usage in Section~\ref{sec:webppl_intro}. All programs can be run using exhaustive enumeration or rejection sampling as inference algorithms, which we elaborate further in Section~\ref{PearlSec}. \\

The three functions can be executed in WebPPL and the posteriors
visualized using the command \lstinline|viz(Infer(prog1))|. The
posterior disease probabilities of each of the programs are
respectively:
\begin{itemize}
\item \texttt{prog1}: $64\%$
\item \texttt{prog2}: $9\%$
\item \texttt{prog3}: $33\%$
\end{itemize}

\noindent The same probabilities appear in the mathematical analysis
in Examples~\ref{PearlDiseaseEx} and~\ref{JeffreyDiseaseEx} below. 

An interesting question to ask is: suppose we do not have 3 tests (2
positive, 1 negative), but 3000 tests (2000 positive, 1000 negative).
Does that change the outcome of the above computations? Not so for the
second and third program, which only require a statistical sample of the data. The first program however, quickly converges to
$100\%$ disease probability when the number of tests increases (still
assuming the same ratio of 2 positive and 1 negative). But this first
program becomes increasingly difficult to compute, because each test result emits further conditioning instructions that the inference engine needs to take into account. The two other programs on the other hand scale almost trivially. We
return to this scaling issue at the end of
Section~\ref{LikelihoodSec}. 

The three implementations will be reiterated throughout the paper and
related to Pearl's and Jeffrey's update. In Section~\ref{BouncerSec},
where we also make their semantics explicit using rejection samplers.

\section{Multisets, Distributions, and Channels}\label{DistributionSec}

Sections~\ref{DistributionSec} -- \ref{JeffreySec} introduce the
mathematics underlying the update situations that we are looking at.
This material is in essence a recap from~\cite{Jacobs19c,Jacobs21c}.
We write $\Mlt$ and $\Dst$ for the multiset and distribution monads on
the category $\Sets$ of sets and functions. For a set $X$, multisets
$\varphi\in\Mlt(X)$ can equivalently be written as a function
$\varphi\colon X \rightarrow \NNO$ with finite support, or as a finite
formal sum $\sum_{i} n_{i}\ket{x_i}$, where $n_{i}\in\NNO$ is the
multiplicity of element $x_{i}\in X$. Similarly, a distribution
$\omega\in\Dst(X)$ is written either as a function $\omega\colon X
\rightarrow [0,1]$ with finite support and $\sum_{x}\omega(x) = 1$, or
as a finite formal convex combination $\sum_{i} r_{i}\ket{x_i}$ with
$r_{i}\in [0,1]$ satisfying $\sum_{i}r_{i} = 1$. 

Functoriality of $\Mlt$ (and $\Dst$) works in the following manner.
For a function $f\colon X \rightarrow Y$ we have $\Mlt(f) \colon
\Mlt(X) \rightarrow \Mlt(Y)$, given as $\Mlt(f)(\varphi)(y) =
\sum_{x\in f^{-1}(y)} \varphi(x)$.

For a multiset $\varphi\in\Mlt(X)$ we write $\|\varphi\| \in \NNO$ for
its size, defined as sum of its multiplicities: $\|\varphi\| \coloneqq
\sum_{x}\varphi(x)$. When this size is not zero, we can define an
associated distribution $\flrn(\varphi) \in \Dst(X)$, via frequentist
learning (normalisation), as:
\[ \begin{array}{rcl}
\flrn(\varphi)
& \coloneqq &
\displaystyle\sum_{x\in X} \frac{\varphi(x)}{\|\varphi\|}\,\bigket{x}.
\end{array} \]

For $K\in\NNO$ we write $\Mlt[K](X) =
\setin{\varphi}{\Mlt(X)}{\|\varphi\|=K}$ for the set of multiset of
size $K$. There is an accumulation function $\acc \colon X^{K}
\rightarrow \Mlt[K](X)$, given by $\acc(x_{1}, \ldots, x_{K}) =
1\ket{x_1} + \cdots + 1\ket{x_K}$. For instance $\acc(a,b,a,c,a,b) =
3\ket{a} + 2\ket{b} + 1\ket{c}$, using $X = \{a,b,c\}$ and $K=6$.

For two distributions $\omega\in\Dst(X)$, $\rho\in\Dst(Y)$ one can
form the (parallel) product distribution $\omega\otimes\rho \in
\Dst(X\times Y)$, with $\big(\omega\otimes\rho\big)(x,y) =
\omega(x)\cdot\rho(y)$. We often use the $K$-fold product $\omega^{K}
= \omega\otimes\cdots\otimes\omega \in \Dst(X^{K})$.

A distribution $\omega\in\Dst(X)$ may be seen as an urn with
coloured balls, where $X$ is the set of colours. The number
$\omega(x)\in [0,1]$ is the probability of drawing a ball of colour
$x$. We are interested in $K$-sized draws, formalised as multiset
$\varphi\in\Mlt[K](X)$. The multinomial distribution
$\multinomial[K](\omega) \in \Dst\big(\Mlt[K](X)\big)$ assigns
probabilities to such draws:
\begin{equation}
\label{MultinomialEqn}
\begin{array}{rclcrcl}
\multinomial[K](\omega)
\hspace*{\arraycolsep}\coloneqq\hspace*{\arraycolsep}
\Dst(\acc)\big(\omega^{K}\big)
& = &
\displaystyle\sum_{\varphi\in\Mlt[K](X)} \, \coefm{\varphi}\cdot
   \prod_{x\in X} \omega(x)^{\varphi(x)}\,\bigket{\varphi}
& \mbox{\quad where \quad} &
\coefm{\varphi}
& \coloneqq &
\displaystyle\frac{\|\varphi\|!}{\prod_{x} \varphi(x)!}.
\end{array}
\end{equation}

A Kleisli map $c\colon X \rightarrow \Dst(Y)$ for the distribution
monad $\Dst$ is often called a \emph{channel}, and written as $c\colon
X \chanto Y$. For instance, the above accumulation map $\acc\colon
X^{K} \rightarrow \Mlt[K](X)$ has a probabilistic inverse $\arr \colon
\Mlt[K](X) \rightarrow \Dst(X^{K})$, where $\arr$ stands for
arrangement, see~\cite{Jacobs21b} for details. This arrangement is
defined as:
\begin{equation}
\label{ArrEqn}
\begin{array}{rcl}
\arr(\varphi)
& \coloneqq &
\displaystyle \sum_{\vec{x}\in\acc^{-1}(\varphi)} \, \frac{1}{\coefm{\varphi}}
   \,\bigket{\vec{x}}
\qquad\mbox{with $\coefm{\varphi}$ as defined in~\eqref{MultinomialEqn}.}
\end{array}
\end{equation}

Kleisli extension gives a pushforward operation along a channel: a
distribution $\omega\in\Dst(X)$ can be turned into a distribution $c
\push \omega \in\Dst(Y)$ via the formula:
\[ \begin{array}{rcl}
c\push \omega
& \coloneqq &
\displaystyle\sum_{y\in Y} \left(\sum_{x\in X} \omega(x)\cdot c(x)(y)\right)
   \bigket{y}.
\end{array} \]

\noindent This new distribution $c \push \omega$ is often called the
prediction. One can prove: $\flrn \push \multinomial[K](\omega) =
\omega$ and $\arr \push \multinomial[K](\omega) = \omega^{K}$,
see~\cite{Jacobs21b}.

The following two programs are equivalent ways of sampling from a prediction $c\push\omega$:
\begin{equation}
	\label{PushProg}
	\vcenter{\hbox{\begin{minipage}[t]{0.3\textwidth}
				\begin{lstlisting}
x <- $\omega$
y <- $c$(x)
samples.add(y)
				\end{lstlisting}
	\end{minipage}}}
	\hspace*{10em}
	\vcenter{\hbox{\begin{minipage}[t]{0.3\textwidth}
				\begin{lstlisting}
y <- $c \push \omega$
samples.add(y)
				\end{lstlisting}
	\end{minipage}}}
\end{equation}
It shows that such sampling can be done in two
steps: The notation $x \leftarrow \omega$ is used for sampling a
random element $x\in X$ from a distribution $\omega\in\Dst(X)$, where
the randomness takes the probabilities in $\omega$ into account. This
is a standard construct in probabilistic programming. If multiple
samples $x_{i} \leftarrow \omega$ are taken, and accumulated in a
multiset $\varphi\in\Mlt(X)$, then the normalisation $\flrn(\varphi)$
of $\varphi$ approaches the original distribution $\omega$. 

Lastly, the tensor product $\otimes$ extends pointwise to channels:
$(c\otimes d)(x,y) = c(x) \otimes d(y)$. Then one can prove, for
instance, $(c \otimes d) \push (\omega\otimes\rho) = (c
\push\omega)\otimes (d\push\rho)$.

\section{Validity, Conditioning, and Pearl's Update Rule}\label{PearlSec}

A (fuzzy) predicate on a set $X$ is a function $p\colon X \rightarrow
[0,1]$.  Each element $x\in X$ gives rise to a \emph{point predicate}
$\indic{x} \colon X \rightarrow [0,1]$, with $\indic{x}(y) = 1$ if
$x=y$ and $\indic{x}(y) = 0$ if $x\neq y$. For two predicates
$p_{1},p_{2} \colon X \rightarrow [0,1]$ we can form a conjunction
$p_{1}\andthen p_{2} \colon X \rightarrow [0,1]$ via pointwise
multiplication: $(p_{1}\andthen p_{2})(x) = p_{1}(x) \cdot p_{2}(x)$.

The validity (or expected value) of a predicate $p\colon X \rightarrow
[0,1]$ in a distribution $\omega\in\Dst(X)$ is written as
$\omega\models p$ and defined as:
\[ \begin{array}{rcl}
\omega\models p
& \,\coloneqq\,
\displaystyle\sum_{x\in X} \omega(x)\cdot p(x).
\end{array} \]

\noindent When this validity is non-zero we can define the updated
distribution $\omega|_{p} \in \Dst(X)$ as:
\begin{equation}
\label{ConditionEqn}
\begin{array}{rcl}
\omega|_{p}
& \coloneqq &
\displaystyle\sum_{x\in X} \frac{\omega(x)\cdot p(x)}{\omega\models p}
   \,\bigket{x}.
\end{array}
\end{equation}

For a channel $c\colon X \chanto Y$ and a predicate $q\colon Y
\rightarrow [0,1]$ on its codomain, we can define a pullback predicate
$c \pull q$ on $X$ via the formula:
\[ \begin{array}{rcl}
\big(c \pull q\big)(x)
& \coloneqq &
\displaystyle\sum_{y\in Y} c(x)(y)\cdot q(y).
\end{array} \]

\noindent The following result contains the basic facts that we need
here. Proofs can be found for instance in~\cite{Jacobs19c,Jacobs21c}.

\begin{lemma}
\label{ValidityLem}
For a channel $c\colon X \chanto Y$, a distribution $\omega\in\Dst(X)$,
predicates $p,p_{1},p_{2}$ on $X$ and $q$ on $Y$,
\begin{enumerate}
\item \label{ValidityLemVal} $c \push \omega \models q \,=\, \omega
  \models c \pull q$;

\item \label{ValidityLemConj} $\omega|_{p_{1}}|_{p_{2}} =
  \omega|_{p_{1} \andthen p_{2}}$;

\item \label{ValidityLemInc} $\omega|_{p} \models p \,\geq\,
  \omega\models p$. \QED
\end{enumerate}
\end{lemma}

The last result shows that a predicate $p$ is `more true' in an
updated distribution $\omega|_{p}$ than in the original $\omega$.  The
next result from~\cite{Jacobs19c,Jacobs21c} contains both the
formulation of Pearl's update, and the associated validity increase.

\begin{theorem}
\label{PearlThm}
Let $c\colon X \chanto Y$ be a channel with a prior distribution
$\omega\in\Dst(X)$ on its domain and a predicate $q\colon Y
\rightarrow [0,1]$ on its codomain. The posterior distribution
$\omega_{P}\in\Dst(X)$ of $\omega$, via Pearl' update rule, with the
evidence predicate $q$, is defined as:
$$\begin{array}{rclcrcl}
\omega_{P}
& \coloneqq &
\omega|_{c \pull q}
& \mbox{\qquad and satisfies \qquad} &
c \push \omega_{P} \models q
& \,\geq\, &
c \push \omega \models q.
\end{array} \eqno{\QEDbox}$$
\end{theorem}

The proof follows from an easy combination of
points~\eqref{ValidityLemVal} and~\eqref{ValidityLemInc} of
Lemma~\ref{ValidityLem}. The increase in validity that is achieved via
Pearl's rule means that the validity of predicate $q$ is higher in the
predicted distribution obtained from the posterior distribution
$\omega_{P}$, than in the prediction obtained from original, prior
distribution $\omega$.

The following are two rejection samplers that allow sampling from a posterior distribution: On the left below we show how to obtain an updated distribution
$\omega|_{p}$ via sampling, and on the right how to get a Pearl
update $\omega|_{c \pull q}$.
\begin{ceqn}
\begin{equation}
\label{UpdateProg}
\vcenter{\hbox{\begin{minipage}[t]{0.3\textwidth}
\begin{lstlisting}
x <- $\omega$
y <- flip($p$(x))
if y == 1:
   samples.add(x)
\end{lstlisting}
\end{minipage}}}
\hspace*{8em}
\vcenter{\hbox{\begin{minipage}[t]{0.3\textwidth}
\begin{lstlisting}
x <- $\omega$
y <- $c$(x)
z <- flip($q$(y))
if z == 1:
   samples.add(x)
\end{lstlisting}
\end{minipage}}}
\end{equation}
\end{ceqn}

\noindent The probabilistic program \texttt{prog1} at the end of
Section~\ref{ExSec} computes the Pearl update. How this update works
in detail will be described next.

\begin{example}
\label{PearlDiseaseEx}
We are now in a situation to explain the $64\%$ posterior disease
probability claimed in Section~\ref{ExSec}. It is obtained via
repeated Pearl updates. We first translate the information given there
into mathematical structure.

We use $X = \{d,\no{d}\}$ for the set with elements $d$ for disease
and $\no{d}$ for no-disease. The given prevalence of $5\%$ for the
disease corresponds to a prior distribution $\omega\in\Dst(X)$ given
by $\omega = \frac{1}{20}\ket{d} + \frac{19}{20}\ket{\no{d}}$.

The test is formalised as a channel $c\colon X \rightarrow \Dst(Y)$
where $Y = \{p,n\}$ the set of positive and negative test
outcomes. The sensitivity and specificity of the test translate into,
respectively:
$$\begin{array}{rclcrcl}
c(d)
& \coloneqq &
\frac{9}{10}\ket{p} + \frac{1}{10}\ket{n}
& \mbox{\qquad and \qquad} &
c(\no{d})
& \coloneqq &
\frac{1}{20}\ket{p} + \frac{19}{20}\ket{n}.
\end{array}$$

\noindent There are two obvious point predicates $\indic{p}\colon
Y\rightarrow [0,1]$ and $\indic{n}\colon Y \rightarrow [0,1]$ on the
set $Y = \{p,n\}$ of test outcomes. We are told that there are two
positive and one negative test. This translates in the conjunction $(c
\pull \indic{p}) \andthen (c \pull \indic{p}) \andthen (c \pull
\indic{n})$. Since conjunction is commutative, the order does not matter.
Updating with this conjection is equivalent to three successive update,
see Lemmma~\ref{ValidityLem}~\eqref{ValidityLemConj}, and gives the
claimed outcome:
\[ \begin{array}{rcl}
\omega_{P}
\hspace*{\arraycolsep}=\hspace*{\arraycolsep}
\omega\big|_{(c \pull \indic{p}) \andthen (c \pull \indic{p}) \andthen (c \pull \indic{n})}
\hspace*{\arraycolsep}=\hspace*{\arraycolsep}
\omega\big|_{c \pull \indic{p}}\big|_{c \pull \indic{p}}\big|_{c \pull \indic{n}}
& = &
\frac{648}{1009}\bigket{d} + \frac{361}{1009}\bigket{\no{d}}
\\
& \approx &
0.642\bigket{d} + 0.358\bigket{\no{d}}.
\end{array} \]

\noindent This is the probability computed in \texttt{prog1}
in Section~\ref{ExSec}.

The validity increase associated with Pearl's update rule
takes the following form.
\[ \begin{array}{rcccccl}
c \push \omega_{P} \models (c \pull \indic{p})^{2} \andthen (c \pull \indic{n})
& \approx &
0.049
& \,\geq\, &
0.0096
& \approx &
c \push \omega \models (c \pull \indic{p})^{2} \andthen (c \pull \indic{n}).
\end{array} \]
\end{example}

\section{Dagger channels and Jeffrey's update rule}\label{JeffreySec}

First we recall that the difference (divergence) between two
distributions $\omega,\rho\in\Dst(X)$ is commonly expressed as
\emph{Kullback-Leibler divergence}, defined as:
\begin{equation}
\label{KLEqn}
\begin{array}{rcl}
\DKL\big(\omega, \, \rho\big)
& \coloneqq &
\displaystyle\sum_{x\in X} \omega(x)\cdot 
   \ln\left(\frac{\omega(x)}{\rho(x)}\right),
  \mbox{\qquad where $\ln$ is the natural logarithm.}
\end{array}
\end{equation}

\noindent The main ingredient that we need for Jeffrey's rule is the
dagger of a channel $c\colon X \chanto Y$ with respect to a prior
distribution $\omega\in\Dst(X)$. This dagger is a channel
$c^{\dag}_{\omega} \colon Y \chanto X$ in the opposite direction. It
is also called Bayesian inversion, see~\cite{ClercDDG17,ChoJ19}, and
it is defined on $y\in Y$ as:
\begin{equation}
\label{DaggerEqn}
\begin{array}{rcccl}
c^{\dag}_{\omega}(y)
& \coloneqq &
\omega|_{c \pull \indic{y}}
& \smash{\stackrel{\eqref{ConditionEqn}}{=}} &
\displaystyle\sum_{x\in X} \, \frac{\omega(x)\cdot c(x)(y)}{(c \push \omega)(y)}
   \,\bigket{x}.
\end{array}
\end{equation}

\noindent We again combine Jeffrey's rule with its main divergence
reduction property, from~\cite{Jacobs21c}. The set-up is very much as
for Pearl's rule, in Theorem~\ref{PearlThm}, but with evidence now in
the form of distribution instead of a predicate.

\begin{theorem}
\label{JeffreyThm}
Let $c\colon X \chanto Y$ be a channel with a prior distribution
$\omega\in\Dst(X)$ and an evidence distribution $\tau\in\Dst(Y)$. The posterior distribution $\omega_{J}\in\Dst(X)$ of $\omega$,
obtained via Jeffrey's update rule, with the evidence distribution $\tau$,
is defined as:
$$\begin{array}{rclcrcl}
\omega_{J}
& \coloneqq &
c^{\dag}_{\omega} \push \tau
& \mbox{\qquad and satisfies \qquad} &
\DKL\big(\tau, \, c \push \omega_{J}\big)
& \,\leq\, &
\DKL\big(\tau, \, c \push \omega\big).
\end{array} \eqno{\QEDbox}$$
\end{theorem}

The proof of this divergence decrease is remarkably hard,
see~\cite{Jacobs21c} for details. The result says that the prediction
from $\omega_{J}$ is less wrong than from $\omega$, when compared to
the `target' distribution $\tau$.



\begin{example}
\label{JeffreyDiseaseEx}
We build on the test channel $c\colon X \chanto Y$ and prevalence
distribution $\omega\in\Dst(X)$ from Example~\ref{PearlDiseaseEx}.
The first task is to compute the dagger channel $f \coloneqq
c^{\dag}_{\omega} \colon Y \chanto X$. It yields:
$$\begin{array}{rclcrcl}
f(p)
& = &
\frac{18}{37}\bigket{d} + \frac{19}{37}\bigket{\no{d}}
& \mbox{\qquad and \qquad} &
f(n)
& = &
\frac{2}{363}\bigket{d} + \frac{361}{363}\bigket{\no{d}}.
\end{array}$$

\noindent The fact that there are two positive and one negative test
translates into the `empirical' evidence distribution $\tau =
\frac{2}{3}\ket{p} + \frac{1}{3}\ket{n} \in \Dst(Y)$. The posterior,
updated disease distribution, obtained from this evidence, gives the
$33\%$ probability mentioned in Section~\ref{ExSec}:
\[ \begin{array}{rcccccl}
\omega_{J}
& = &
f \push \tau
& = &
\frac{13142}{40293}\bigket{d} + \frac{27151}{40293}\bigket{\no{d}}
& \approx &
0.326\bigket{d} + 0.674\bigket{\no{d}}.
\end{array} \]

\noindent This probability is computed by \texttt{prog3} in
Section~\ref{ExSec}.

The divergence decrease from Theorem~\ref{JeffreyThm}
takes the following form:
\[ \begin{array}{rcccccl}
\DKL\big(\tau, \, c \push \omega_{J}\big)
& \approx &
0.24
& \,\leq\, &
0.98
& \approx &
\DKL\big(\tau, \, c \push \omega\big).
\end{array} \]

Having seen this, we may ask: why not use the evidence distribution
$\tau = \frac{2}{3}\ket{p} + \frac{1}{3}\ket{n}$ not as a predicate
$q = \frac{2}{3}\indic{p} + \frac{1}{3}\indic{n}$, and then do a single
Pearl update:
\begin{equation}
\label{SinglePearlEqn}
\begin{array}{rcccl}
\omega|_{c \pull q}
& = &
\frac{2}{23}\ket{d} + \frac{21}{23}\ket{\no{d}}
& \approx &
0.087\ket{d} + 0.913\ket{\no{d}}.
\end{array}
\end{equation}

\noindent This is the distribution computed by program \textbf{prog2}
in Section~\ref{ExSec}.
\end{example}

For future use we record the following standard properties of the
dagger of a channel~\eqref{DaggerEqn}.

\begin{lemma}
\label{DaggerLem}
\begin{enumerate}
\item \label{DaggerLemSeq} Daggers preserve sequential composition:
  for two successive channels $X \stackrel{c}{\chanto} Y
  \stackrel{d}{\chanto} Z$ and a distribution $\omega\in\Dst(X)$,
\[ \begin{array}{rcl}
\big(d \klafter c\big)_{\omega}^{\dag}
& = &
c_{\omega}^{\dag} \klafter d_{c \push \omega}^{\dag}.
\end{array} \]

\item \label{DaggerLemPar} Daggers preserve parallel composition: for
  two channels $c\colon X \chanto A$, $d\colon Y\chanto B$ with
  distributions $\omega\in\Dst(X)$, $\rho\in\Dst(Y$),
\[ \begin{array}{rcl}
\big(c\otimes d\big)_{\omega\otimes\rho}^{\dag}
& = &
c_{\omega}^{\dag} \otimes d_{\rho}^{\dag}.
\end{array} \]
\end{enumerate}
\end{lemma}


\section{An Operational Understanding of Jeffrey's Update}\label{BouncerSec}
\lstset{language=Python}


\begin{wrapfigure}{r}{0.20\textwidth}
	\begin{center}
		\includegraphics[scale=0.5]{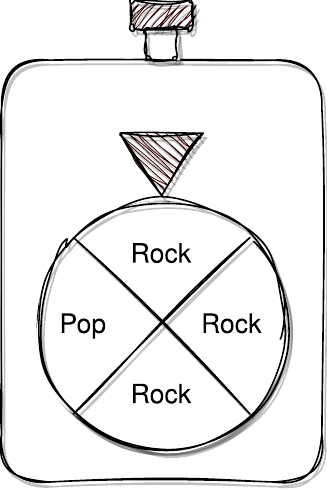}
	\end{center} 
	\label{fig:ticker}
	\caption{Ticker device}
\end{wrapfigure}

\vspace{-0.35cm}

We return to the probabilistic programs of Section~\ref{ExSec}. As
discussed in Section~\ref{PearlSec}, \texttt{prog1} expresses repeated
Pearl updates. It remains to understand the difference between
\texttt{prog2} and \texttt{prog3}. As shown in~\eqref{SinglePearlEqn},
\texttt{prog2} corresponds to a single Pearl's update with the target
distribution, as predicate. Further, \texttt{prog3} is Jeffrey's
update, with the nested inference corresponding to the computation of
the dagger channel $c^\dagger_\omega$. The difference between the two
programs \texttt{prog2} and \texttt{prog3} is surprisingly subtle, so
we begin by illustrating it using a different kind of metaphor, and
derive a rejection sampler for each case in turn.

Consider a large queue of people waiting in front of a club. Each
person prefers either rock or pop. The club's management wants to
achieve a target ratio of 75\% rock fans on the inside. To that end,
they equip their doorman with a special ticker device, see
Figure~\ref{fig:ticker}. The ticker displays a current target (either
`Rock' or `Pop'), and the doorman admits the next person if and only
if they prefer the targeted style. The doorman can click the device to
obtain a new target (either by cycling sequentially through the
targets, or picking one randomly), but there remains a choice when to
click.
\begin{enumerate}
\item Single Pearl Policy: pick a new target after every person:
\begin{lstlisting}
for person in queue:
  if person.preference == ticker.target:
    club.admit(person)
  ticker.click()
\end{lstlisting}
\item Jeffrey Policy: pick a new target only after admitting a person:
\begin{lstlisting}
for person in queue:
  if person.preference == ticker.target:
    club.admit(person)
    ticker.click()
\end{lstlisting}
\end{enumerate} 

\noindent It may be clear that only the Jeffrey Policy is suitable to
achieve the management's goal. Approximately 75\% of the people which
are admitted are rock fans. This is in line with the key property of
Jeffrey's update rule: reducing the divergence with the target
distribution $\tau$, see Theorem~\ref{JeffreyThm}. It is unclear what
the single Pearl policy achieves in this context.

We may also wonder how the door policy influences other statistical properties of the audience (such as age or gender) which may correlate with music preference: If the prior distribution in the queue is $\omega$, what will the resulting distribution be inside the club?  For the Jeffrey Policy, this update is precisely described by Jeffrey's update. We summarize this section with a concrete description of rejection samplers for Pearl's update with a random target (left) and Jeffrey's update (right), corresponding to the semantics of the probabilistic programs \texttt{prog2} and \texttt{prog3}:

\begin{ceqn}
	\begin{equation}
		\vcenter{\hbox{\begin{minipage}[t]{0.3\textwidth}
					\begin{lstlisting}
while True:
  x <- $\omega$
  y <- c(x)
  target <- $\tau$
  if y == target:
    samples.add(x)
					\end{lstlisting}
		\end{minipage}}}
		\hspace*{5em}
		\vcenter{\hbox{\begin{minipage}[t]{0.3\textwidth}
					\begin{lstlisting}
while True:
  x <- $\omega$
  y <- c(x)
  if y == target:
    samples.add(x)
    target <- $\tau$
					\end{lstlisting}
		\end{minipage}}}
	\end{equation}
\end{ceqn}

\section{Likelihoods and Generative Models for Pearl and Jeffrey}\label{LikelihoodSec}

This section first identifies two forms of likelihood of data in the
situation with a statistical model given by a channel $X\chanto Y$ and
a distribution on $X$.  It then relates these two forms of likelihood
to the two update rules of repeated-Pearl and Jeffrey --- in
Theorems~\ref{PearlThm} and~\ref{JeffreyThm}.

\begin{definition}
\label{LikelihoodDef}
Let $\psi\in\Mlt[K](Y)$ be a multiset of data, of size $K =
\|\psi\|\in\NNO$. Let $c\colon X \chanto Y$ be a channel with a
distribution $\omega\in\Dst(X)$ on its domain. 
\begin{enumerate}
\item \label{LikelihoodDefJeffrey} The \emph{Jeffrey likelihood} of
  the multiset $\psi$ is given by the number:
\[ \multinomial[K]\big(c \push \omega\big)(\psi). \]

\item \label{LikelihoodDefPearl} The \emph{Pearl likelihood} of $\psi$ in
the same model is the first expression below, which has several alternative
formulations. It uses the abbreviation $\multinomial[K](c) \coloneqq
\multinomial[K] \after c$.
\[ \begin{array}{rcl}
\big(\multinomial[K](c) \push \omega\big)(\psi)
& = &
\multinomial[K](c) \push \omega \models \indic{\psi}
\\
& = &
\omega \models \multinomial[K](c) \pull \indic{\psi}
   \qquad\mbox{by Lemma~\ref{ValidityLem}~\eqref{ValidityLemVal}.}
\end{array} \]
\end{enumerate}
\end{definition}


\begin{figure}
\[ \input{jeffrey-pearl.tikz} \]
\caption{Graphical representation of Jeffrey likelihood on the left,
  and Pearl likelihood on the right, see
  Definition~\ref{LikelihoodDef}.}
\label{PlatesFig}
\end{figure}
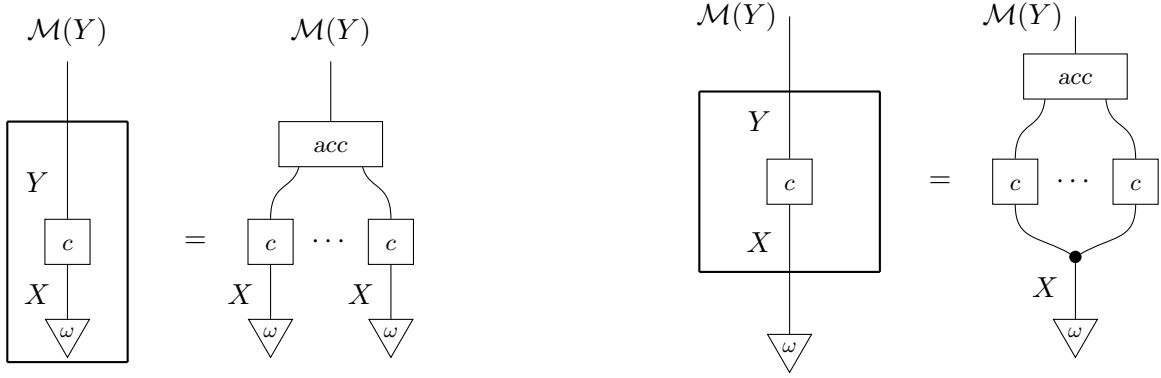

Associated to these two likelihoods are different generative models,
\textit{i.e.}~distributions over multisets, in $\Dst(\Mlt[K](Y))$,
which we evaluate on the dataset $\psi$. For Jeffrey likelihood in
item~\eqref{LikelihoodDefJeffrey} we first do the Kleisli extension
$c\push(\cdot)$ of $c$ and then take the multinomial, as in the
composite:
\[ \xymatrix{
	\Dst(X)\ar[rr]^-{c\push(\cdot)} & & \Dst(Y)\ar[rr]^-{\multinomial[K]}
	& & \Dst\Big(\Mlt[K](Y)\Big).
} \]

\noindent We can concisely illustrate this with string diagrams using
an informal `plate' notation to copy parts of the string diagram
(inspired by the use of plates in graphical models), see
Figure~\ref{PlatesFig} on the left. In contrast, for the Pearl
likelihood in item~\eqref{LikelihoodDefPearl} we use the composite
$\multinomial[K](c) \coloneqq \multinomial[K] \after c$ in the
pushforward:
\[ \xymatrix@C+1.0pc{
\Dst(X)\ar[rr]\ar[rr]^-{\multinomial[K](c)\push(\cdot)}
   & & \Dst\Big(\Mlt[K](Y)\Big).
} \]

\noindent Here, the plate does not extend over the distribution
$\omega$, whose output is copied instead of resampled, see
Figure~\ref{PlatesFig} on the right.

The Pearl likelihood is used in the multinomial naive Bayes
classifier~\cite{naivebayes}. For the likelihood of Jeffrey we shall
see alternative formulations in Section~\ref{JeffreyVISec} below.

Our first result says that minimising the Kullback-Leibler divergence
that occurs in Theorem~\ref{JeffreyThm} --- and that is actually
reduced by Jeffrey's update rule --- corresponds to maximising the
Jeffrey likelihood of
Definition~\ref{LikelihoodDef}~\eqref{LikelihoodDefJeffrey}.

\begin{theorem}
\label{JeffreyLikelihoodThm}
\begin{enumerate}
\item For distributions $\omega,\omega'\in\Dst(X)$ and channels
  $c,c'\colon X \chanto Y$, with data $\psi\in\Mlt(Y)$, we have that
  Jeffrey likelihood is oppositely ordered to Kullback-Leibler
  divergence in:
\[ \begin{array}{rcl}
\multinomial[K]\big(c \push \omega\big)(\psi)
  \leq \multinomial[K]\big(c' \push \omega'\big)(\psi)
& \Longleftrightarrow &
\DKL\big(\flrn(\psi), \, c \push \omega\big)
  \geq \DKL\big(\flrn(\psi), \, c' \push \omega'\big).
\end{array} \]

\item Fix a channel $c\colon X \chanto Y$. Then:
\[ \begin{array}{rcl}
\argmax\limits_{\omega\in\Dst(X)} \, \multinomial[K]\big(c \push \omega\big)(\psi)
& = &
\argmin\limits_{\omega\in\Dst(X)} \, 
   \DKL\big(\flrn(\psi), \, c \push \omega\big).
\end{array} \]
\end{enumerate}
\end{theorem}

The above expression on the right is the divergence between the data
distribution and the prediction $c \push \omega$. This divergence can
be reduced via Jeffrey's rule. The above result says that Jeffrey's
rule thus increases the Jeffrey likelihood, see
Theorem~\ref{JeffreyThm}.

\begin{myproof}
We only prove the first item, since the second one is a direct
consequence.  We use that the natural logarithm $\ln \colon \pR
\rightarrow \R$ preserves and reflects the order: $a \leq b$ iff
$\ln(a) \leq \ln(b)$. This is used in the first step below. We
additionally use that the logarithm sends multiplications to sums.
\[ \begin{array}[b]{rcl}
\lefteqn{\multinomial[K]\big(c \push \omega\big)(\psi)
  \leq \multinomial[K]\big(c' \push \omega'\big)(\psi)}
\\[+0.3em]
& \Longleftrightarrow &
\ln\Big(\multinomial[K]\big(c \push \omega\big)(\psi)\Big)
  \leq \ln\Big(\multinomial[K]\big(c' \push \omega'\big)(\psi)\Big)
\\[+0.6em]
& \Longleftrightarrow &
\displaystyle
\ln\left(\coefm{\psi} \cdot \prod_{y\in Y}\, (c\push\omega)(y)^{\psi(y)}\right)
   \leq
\ln\left(\coefm{\psi} \cdot \prod_{y\in Y}\, (c'\push\omega')(y)^{\psi(y)}\right)
\\[+1.4em]
& \Longleftrightarrow &
\displaystyle
\ln\Big(\coefm{\psi}\Big) \,+\,
   \sum_{y\in Y}\, \psi(y) \cdot \ln\Big((c\push\omega)(y)\Big)
   \leq
\ln\Big(\coefm{\psi}\Big) \,+\,
   \sum_{y\in Y}\, \psi(y) \cdot \ln\Big((c'\push\omega')(y)\Big)
\\
& \Longleftrightarrow &
	\displaystyle
	- \sum_{y\in Y}\, \frac{\psi(y)}{\|\psi\|} \cdot 
	\ln\Big((c\push\omega)(y)\Big)
	\geq 
	- \sum_{y\in Y}\, \frac{\psi(y)}{\|\psi\|} \cdot 
	\ln\Big((c'\push\omega')(y)\Big)
	\\[+1.2em]
	& \Longleftrightarrow &
	\displaystyle
	\sum_{y\in Y}\, \flrn(\psi)(y) \cdot \ln\Big(\flrn(\psi)(y)\Big) 
	- \sum_{y\in Y}\, \flrn(\psi)(y) \cdot 
	\ln\Big((c\push\omega)(y)\Big)
	\\[+1.2em]
	& & \hspace*{5em} \displaystyle \geq\,
	\sum_{y\in Y}\, \flrn(\psi)(y) \cdot \ln\Big(\flrn(\psi)(y)\Big)
	- \sum_{y\in Y}\, \flrn(\psi)(y) \cdot 
	\ln\Big((c'\push\omega')(y)\Big)
	\\[+1.2em]
	& \Longleftrightarrow &
	\displaystyle
	\sum_{y\in Y}\, \flrn(\psi)(y) \cdot 
	\ln\left(\frac{\flrn(\psi)(y)}{(c\push\omega)(y)}\right)
	\geq 
	\sum_{y\in Y}\, \flrn(\psi)(y) \cdot 
	\ln\left(\frac{\flrn(\psi)(y)}{(c'\push\omega')(y)}\right)
	\\[+1.2em]
	& \Longleftrightarrow &
	\DKL\big(\flrn(\psi), \, c \push \omega\big)
	\geq \DKL\big(\flrn(\psi), \, c' \push \omega'\big).
\end{array} \eqno{\QEDbox} \]

\auxproof{
\[ \begin{array}[b]{rcl}
\lefteqn{\argmax\limits_{(\omega,c)\in M} \, 
   \multinomial[K]\big(c \push \omega\big)(\psi)}
\\
& = &
\argmax\limits_{(\omega,c)\in M} \, 
   \ln\Big(\multinomial[K]\big(c \push \omega\big)(\psi)\Big)
\\
& = &
\displaystyle\argmax\limits_{(\omega,c)\in M} \, 
   \ln\left(\coefm{\psi} \cdot \prod_{x\in X}\, (c\push\omega)(x)^{\psi(x)}\right)
\\[+1em]
& = &
\displaystyle\argmax\limits_{(\omega,c)\in M} \, 
   \ln\Big(\coefm{\psi}\Big) \,+\,
   \sum_{x\in X}\, \psi(x) \cdot \ln\Big((c\push\omega)(x)\Big)
\\[+1.2em]
& = &
\displaystyle\argmax\limits_{(\omega,c)\in M} \, \|\psi\| \cdot
   \sum_{x\in X}\, \frac{\psi(x)}{\|\psi\|} \cdot 
   \ln\Big((c\push\omega)(x)\Big) \,-\,
   \sum_{x\in X}\, \flrn(\psi)(x) \cdot \ln\Big(\flrn(\psi)(x)\Big)
\\[+1em]
& = &
\displaystyle\argmax\limits_{(\omega,c)\in M} \, 
   - \sum_{x\in X}\, \flrn(\psi)(x) \cdot 
   \ln\left(\frac{\flrn(\psi)(x)}{(c\push\omega)(x)}\right)
\\[+1em]
& = &
\argmin\limits_{(\omega,c)\in M} \, 
   \DKL\big(\flrn(\psi), \, c \push \omega\big).
\end{array} \eqno{\QEDbox} \]
}
\end{myproof}

We also relate Pearl likelihood to Pearl's update rule.

\begin{theorem}
\label{PearlLikelihoodThm}
Consider a channel $c\colon X\chanto Y$ with distribution
$\omega\in\Dst(X)$ and data $\psi\in\Mlt[K](Y)$. The validity increase
of Theorem~\ref{PearlThm}, applied to the last formulation of Pearl
likelihood in
Definition~\ref{LikelihoodDef}~\eqref{LikelihoodDefPearl}, gives an
increase of Pearl likelihood via a repetition of Pearl's rule:
\[ \begin{array}{rcl}
\big(\multinomial[K](c) \push \omega\big)(\psi)
& \,=\, &
\omega \models \multinomial[K](c) \pull \indic{\psi}
\\
& \,\leq\, &
\omega|_{\multinomial[K](c) \pull \indic{\psi}} \models 
   \multinomial[K](c) \pull \indic{\psi}
\hspace*{\arraycolsep}\,=\,\hspace*{\arraycolsep}
\big(\multinomial[K](c) \push 
   \omega|_{\multinomial[K](c) \pull \indic{\psi}}\big)(\psi).
\end{array} \]

\noindent This updated distribution $\omega|_{\multinomial[K](c) \pull
  \indic{\psi}} = \multinomial[K](c)^{\dag}_{\omega}(\psi) \in
\Dst(Y)$ can be described via repeated Pearl updates as:
\[ \begin{array}{rcl}
\omega|_{\multinomial[K](c) \pull \indic{\psi}}
& = &
\omega|_{\andthen_{y\in Y} (c \pull \indic{y})^{\psi(y)}}
\\
& = &
\omega|_{(c \pull \indic{y_1})^{\psi(y_{1})} \,\andthen\, \cdots \,\andthen\, 
   (c \pull \indic{y_n})^{\psi(y_{n})}}
   \qquad \mbox{if } \supp(\psi) = \{y_{1}, \ldots, y_{n}\}
\\
& = &
\omega|_{\underbrace{\scriptstyle(c \pull \indic{y_1}) 
   \,\andthen\, \cdots \,\andthen\, 
   (c \pull \indic{y_1})}_{\psi(y_{1})\text{ times}}
   \;\andthen\; \cdots \;\andthen\;
   \underbrace{\scriptstyle(c \pull \indic{y_n}) 
   \,\andthen\, \cdots \,\andthen\, 
   (c \pull \indic{y_n})}_{\psi(y_{n})\text{ times}}}
\\
& = &
\omega|_{c \pull \indic{y_1}} \!\cdots |_{c \pull \indic{y_1}} \cdots\,
   |_{c \pull \indic{y_n}} \!\cdots |_{c \pull \indic{y_n}}.
\end{array} \]
\end{theorem}

We have used such successive updates in the calculation of the disease
probabilities according to Pearl in Example~\ref{PearlDiseaseEx}.

\begin{myproof}
We first note that we can write Pearl's likelihood as:
\[ \begin{array}{rcl}
\omega \models \multinomial[K](c) \pull \indic{\psi}
\hspace*{\arraycolsep}=\hspace*{\arraycolsep}
\displaystyle\sum_{x\in X}\, \omega(x)\cdot \multinomial[K]\big(c(x)\big)(\psi)
& = &
\displaystyle\sum_{x\in X}\, \omega(x)\cdot \coefm{\psi} \cdot
   \prod_{y\in Y} c(x)(y)^{\psi(y)}
\\[+1.2em]
& = &
\coefm{\psi} \cdot\displaystyle\sum_{x\in X}\, \omega(x)\cdot 
   \prod_{y\in Y} (c \pull \indic{y})(x)^{\psi(y)}
\\[+1.2em]
& = &
\coefm{\psi} \cdot\displaystyle\sum_{x\in X}\, \omega(x)\cdot 
   \left(\bigandthen_{y\in Y} (c \pull \indic{y})^{\psi(y)}\right)(x)
\\[+1.2em]
& = &
\displaystyle\coefm{\psi} \cdot \left(\omega\models 
   \bigandthen_{y\in Y} (c \pull \indic{y})^{\psi(y)}\right).
\end{array} \]

\noindent Now, for $x\in X$,
\[ \begin{array}[b]{rcl}
\omega|_{\multinomial[K](c) \pull \indic{\psi}}(x)
\hspace*{\arraycolsep}\smash{\stackrel{\eqref{ConditionEqn}}{=}}\hspace*{\arraycolsep}
\displaystyle\frac{\omega(x) \cdot \multinomial[K]\big(c(x)\big)(\psi)}
   {\omega \models \multinomial[K](c) \pull \indic{\psi}}
& = &
\displaystyle\frac{\omega(x) \cdot \coefm{\psi} \cdot 
   (\andthen_{y} (c \pull \indic{y})^{\psi(y)})(x)}
   {\coefm{\psi} \cdot (\omega\models \andthen_{y} (c \pull \indic{y})^{\psi(y)})}
\\[+1em]
& = &
\displaystyle\frac{\omega(x) \cdot 
   (\andthen_{y} (c \pull \indic{y})^{\psi(y)})(x)}
   {\omega\models \andthen_{y} (c \pull \indic{y})^{\psi(y)}}
\hspace*{\arraycolsep}=\hspace*{\arraycolsep}
\omega|_{\andthen_{y} (c \pull \indic{y})^{\psi(y)}}(x).
\end{array} \eqno{\QEDbox} \]
\end{myproof}

The conjunction predicate $\andthen_{y\in Y} (c \pull
\indic{y})^{\psi(y)}$ used in the above
Theorem~\ref{PearlLikelihoodThm} looses its value in practice as soon
as we have much data, that is, when the multiset $\psi$ is big. The
conjunction involves multiplication of probabilities and thus quickly
becomes unmanageably small. Thus, Pearl update works only (in
practice) for small amounts of data.

There is an exception however, which is beyond the scope of the
current paper. When there is a conjugate prior situation, Pearl
updates may happen via updates of the hyperparameters. This does scale
to big multisets of data.

\section{Jeffrey's Update Rule via Variational Inference}\label{JeffreyVISec}

In this section we like to make the idea precise that Jeffrey's update
rule involves a `population' perspective, in contrast to the individual
perspective in Pearl's rule. We show how Jeffrey's rule emerges 
from updating a multinomial distribution $\multinomial[K](\omega)$.
There are two challenges.
\begin{itemize}
\item A multinomial distribution $\multinomial[K](\omega)$ is a
  distribution on multisets $\Mlt[K](X)$ of size $K$, when
  $\omega\in\Dst(X)$. When we wish to update along a channel $c\colon
  X \chanto Y$ we first have to extend $c$ to a channel $\Mlt[K](c)
  \colon \Mlt[K](X) \chanto \Mlt[K](Y)$. This can be done via an
  extension of the multiset functor to the Kleisli category
  $\Kl(\Dst)$ of the distribution monad $\Dst$. This will occupy us
  first in this section.

Once we have this channel extension $\Mlt[K](c)$, for a multiset of
data $\psi\in\Mlt[K](Y)$ we can form the following update of the
multinomial distribution, abbreviated as $\sigma \in
\Dst\big(\Mlt[K](X)\big)$.
\begin{equation}
\label{MultinomialUpdateAbr}
\begin{array}{rcl}
\sigma
& \coloneqq &
\multinomial[K](\omega)\big|_{\Mlt[K](c) \pull \indic{\psi}}
\end{array}
\end{equation}

\noindent We like to think of this $\sigma$ as a distribution of the
form $\multinomial[K](\omega')$. The obvious way to obtain this
distribution $\omega'$ is via frequentist learning, as $\flrn \push
\sigma$. Indeed, as we have seen before~\eqref{PushProg}, $\flrn \push
\multinomial[K](\rho) = \rho$.  The first of our two main results in
this section is Theorem~\ref{JeffreyFlrnThm}; it says that $\flrn
\push \sigma$ is the Jeffrey update $c^{\dag}_{\omega} \push
\flrn(\psi)$. This is a technically non-trivial result.

\item Next we use techniques from variational
  inference~\cite{mackay2003information,Murphy12}: we like to
  determine the `best' distribution $\omega'$ such that
  $\multinomial[K](\omega')$ approximates the above distribution
  $\sigma$ in~\eqref{MultinomialUpdateAbr}. We thus look for the
  distribution with minimal Kullback-Leibler divergence. There again
  we find Jeffrey's update:
\[ \begin{array}{rcl}
\argmin\limits_{\omega'\in\Dst(X)} \, 
   \DKL\big(\multinomial[K](\omega'),\, \sigma\big)
& = &
c^{\dag}_{\omega} \push \flrn(\psi).
\end{array} \]

\noindent This is the content of our second main result below, 
Theorem~\ref{JeffreyArgminThm}.
\end{itemize}

\subsection{Jeffrey's rule via Frequentist Learning}\label{JeffreyFlrnSubsec}

Taking multisets of a particular size $K\in\NNO$ forms a functor
$\Mlt[K] \colon \Sets \rightarrow \Sets$. This functor can be extended
to the Kleisli category $\Kl(\Dst)$ of the distribution monad $\Dst$.
This works via a distributive law $\Mlt[K]\Dst \Rightarrow
\Dst\Mlt[K]$, see~\cite{DashS20,Jacobs21b}. The extension can also be
written via accumulation and arrangement, see
Lemma~\ref{ExtensionLem}~\eqref{ExtensionLemAccArr} below. We shall
use it in that form.

The resulting extension is still written as $\Mlt[K] \colon \Kl(\Dst)
\rightarrow \Kl(\Dst)$. It sends a set/object $X$ in $\Kl(\Dst)$ to
the set $\Mlt[K](X)$ of mulitsets of size $K$. On a channel/morphism
$c\colon X\chanto Y$ one defines a channel $\Mlt[K](c) \colon
\Mlt[K](X) \chanto \Mlt[K](Y)$ via the distributive law as:
\[ \xymatrix{
\Mlt[K](c) \;\coloneqq\; \Big(\Mlt[K](X)\ar[rr]^-{\Mlt(c)} & &
   \Mlt[K]\big(\Dst(Y)\big)\ar[r]^-{\text{law}} & \Dst\big(\Mlt[K](Y)\big)\Big).
} \]

\noindent Notice that we have written $\Mlt(c)$ for the application of
the multiset functor $\Mlt\colon\Sets\rightarrow\Sets$, in order to
distinguish it from the extension $\Mlt[K] \colon \Kl(\Dst)
\rightarrow\Kl(\Dst)$.

\begin{lemma}
\label{ExtensionLem}
\begin{enumerate}
\item \label{ExtensionLemAccArr} For a channel $c\colon X \chanto Y$
  and a number $K\in\NNO$ the following diagram commutes.
\[ \xymatrix@C+1pc{
X^{K}\ar[rr]|-{\circ}^-{c^K} & & Y^{K}\ar[d]|-{\circ}^{\acc}
\\
\Mlt[K](X)\ar[u]|-{\circ}^{\arr}\ar[rr]|-{\circ}^-{\Mlt[K](c)} & & \Mlt[K](Y)
} \]

\item \label{ExtensionLemNat} Accumulation $\acc$ and frequentist
  learning $\flrn$ are natural transformations between functors
  extended to Kleisli categories:
\[ \xymatrix@C+0.5pc{
\Kl(\Dst)\ar@/^2.0ex/[rr]^-{(-)^K}_-{\big\Downarrow\rlap{$\scriptstyle\acc$}}
   \ar@/_2.0ex/[rr]_-{\Mlt[K]}
   & & \Kl(\Dst)
& 
\Kl(\Dst)\ar@/^2.0ex/[rr]^-{\overline{\Dst}}_-{\big\Downarrow\rlap{$\scriptstyle\multinomial[K]$}}
   \ar@/_2.0ex/[rr]_-{\Mlt[K]}
   & & \Kl(\Dst)
& 
\Kl(\Dst)\ar@/^2.0ex/[rr]^-{\Mlt[K+1]}\ar@/_2.0ex/[rr]_-{\idmap} &
  {\big\Downarrow}\rlap{$\scriptstyle\flrn$} & \Kl(\Dst)
} \]

\noindent The functor $(-)^{K} \colon \Kl(\Dst) \rightarrow \Kl(\Dst)$
is the $K$-fold tensor product, and $\overline{\Dst} \colon \Kl(\Dst)
\rightarrow \Kl(\Dst)$ is the standard extension of a monad to its
Kleisli category, given on $c\colon X\chanto Y$ by $\overline{\Dst}(c)
= \eta \after c \colon \Dst(c) \colon \Dst(X) \chanto \Dst(Y)$, where
$\eta$ is the unit of the monad $\Dst$.
\end{enumerate}
\end{lemma}

\begin{myproof}
This follow from the results in~\cite{Jacobs21b}. \QED
\end{myproof}

A crucial observation is that the formulation of the extension
$\Mlt[K](c)$ in Lemma~\ref{ExtensionLem}~\eqref{ExtensionLemAccArr}
also works for daggers. It demonstrates that `multisets' and `daggers'
commute, see~\eqref{ExtMltDagCommuteEqn} below.

\begin{proposition}
\label{DaggerOfExtensionProp}
Consider a channel $c\colon X \chanto Y$ with a distribution $\omega\in\Dst(X)$
and a number $K\in\NNO$. Then the following diagram of daggers commutes.
\[ \xymatrix@C+1pc{
X^{K}\ar[d]|-{\circ}_{\acc} & & Y^{K}
   \ar[ll]|-{\circ}_-{\big(c^{\dag}_{\omega}\big)^{K}}
\\
\Mlt[K](X) & & \Mlt[K](Y)\ar[ll]|{\circ}_-{\Mlt[K](c)^{\dag}_{\multinomial[K](\omega)}}\ar[u]|-{\circ}_{\arr}
} \]

\noindent This means that the extended multiset functor
$\Mlt[K]\colon\Kl(\Dst) \rightarrow\Kl(\Dst)$ commutes with daggers,
where the original prior distribution $\omega$ is replaced by the
multinomial distribution $\multinomial[K](\omega)$, that is:
\begin{equation}
\label{ExtMltDagCommuteEqn}
\begin{array}{rcl}
\Mlt[K]\big(c_{\omega}^{\dag}\big)
& = &
\Mlt[K](c)^{\dag}_{\multinomial[K](\omega)}.
\end{array}
\end{equation}
\end{proposition}

\begin{myproof}
We concentrate on proving commutation of the diagram, since it
implies~\eqref{ExtMltDagCommuteEqn} via
Lemma~\ref{ExtensionLem}~\eqref{ExtensionLemAccArr}.  We use
Lemma~\ref{DaggerLem}~\eqref{DaggerLemSeq} as first step in:
\[ \begin{array}{rcl}
\Mlt[K](c)_{\multinomial[K](\omega)}^{\dag}
& = &
\Big(\acc \klafter c^{K} \klafter \arr\Big)_{\multinomial[K](\omega)}^{\dag}
\\[+0.6em]
& = &
\arr_{\multinomial[K](\omega)}^{\dag} \klafter 
   \big(c^{K}\big)_{\arr \push \multinomial[K](\omega)}^{\dag} \klafter 
   \acc^{\dag}_{c^{K} \push (\arr \push \multinomial[K](\omega))}
\\[+0.6em]
& = &
\acc \klafter \big(c_{\omega}^{\dag}\big)^{K} \klafter \arr.
\end{array} \]

\noindent This last equation is justified by the three following
steps.
\begin{itemize}
\item The dagger channel $\arr_{\multinomial[K](\omega)}^{\dag} \colon X^{K}
\chanto \Mlt[K](X)$ is determined on $\vec{x}\in X^{K}$ as:
\[ \begin{array}{rcl}
\arr_{\multinomial[K](\omega)}^{\dag}(\vec{x})
& \smash{\stackrel{\eqref{DaggerEqn}}{=}} &
\displaystyle\sum_{\varphi\in\Mlt[K](X)} \,
   \frac{\multinomial[K](\omega)(\varphi)\cdot\arr(\varphi)(\vec{x})}
      {(\arr \push \multinomial[K](\omega))(\vec{x})}\,\bigket{\varphi}
\\[+1.2em]
& = &
\displaystyle\frac{\multinomial[K](\omega)(\acc(\vec{x}))\cdot
   \frac{1}{\coefm{\varphi}}}{\omega^{K}(\vec{x})}\,\bigket{\acc(\vec{x})}
\hspace*{\arraycolsep}=\hspace*{\arraycolsep}
\displaystyle\frac{\prod_{y} \omega(y)^{\acc(\vec{x})(y)}}
   {\prod_{i} \omega(x_{i})}\,\bigket{\acc(\vec{x})}
\hspace*{\arraycolsep}=\hspace*{\arraycolsep}
1\bigket{\acc(\vec{x})}.
\end{array} \]

\item We again use $\arr \push \multinomial[K](\omega) = \omega^{K}$,
  so that we can apply Lemma~\ref{DaggerLem}~\eqref{DaggerLemPar}:
\[ \begin{array}{rcccl}
\big(c^{K}\big)_{\arr \push \multinomial[K](\omega)}^{\dag}
& = &
\big(c^{K}\big)_{\omega^{K}}^{\dag}
& = &
\big(c_{\omega}^{\dag}\big)^{K}.
\end{array} \]

\item For the channel $\acc^{\dag}_{c^{K} \push (\arr \push
  \multinomial[K](\omega))} \colon \Mlt[K](Y) \chanto Y^{K}$ 
we observe that $c^{K} \push (\arr \push \multinomial[K](\omega))
= c^{K} \push \omega^{K} = (c\push \omega)^{K}$ so that:
\[ \begin{array}[b]{rcl}
\acc^{\dag}_{c^{K} \push (\arr \push \multinomial[K](\omega))}(\psi)
\hspace*{\arraycolsep}=\hspace*{\arraycolsep}
\acc^{\dag}_{(c\push \omega)^{K}}(\psi)
& \smash{\stackrel{\eqref{DaggerEqn}}{=}} &
\displaystyle\sum_{\vec{y}\in X^{K}}\,
   \frac{(c\push \omega)^{K}(\vec{y})\cdot \acc(\vec{y})(\psi)}
   {\big(\acc \push (c\push \omega)^{K}\big)(\psi)}\,\bigket{\vec{y}}
\\[+1.0em]
& = &
\displaystyle\sum_{\vec{y}\in \acc^{-1}(\psi)}\,
   \frac{(c\push \omega)^{K}(\vec{y})}
   {\multinomial[K](c\push \omega)(\psi)}\,\bigket{\vec{y}}
\\[+1.0em]
& = &
\displaystyle\sum_{\vec{y}\in \acc^{-1}(\psi)}\, \frac{1}{\coefm{\psi}}
   \,\bigket{\vec{y}}
\hspace*{\arraycolsep}=\hspace*{\arraycolsep}
\arr(\psi).
\end{array} \eqno{\QEDbox} \]
\end{itemize}
\end{myproof}

At this stage we return to Jeffrey likelihood $\multinomial[K]\big(c
\push \omega\big)(\psi)$, as described in
Definition~\ref{LikelihoodDef}~\eqref{LikelihoodDefJeffrey}. Using the
extended functor $\Mlt[K] \colon \Kl(\Dst) \rightarrow \Kl(\Dst)$ and
the fact that multinomial is a natural transformation $\multinomial[K]
\colon \overline{\Dst} \Rightarrow \Mlt[K]$, see
Lemma~\ref{ExtensionLem}~\eqref{ExtensionLemNat}, we get:
\[ \begin{array}{rcl}
\multinomial[K]\big(c \push \omega\big)(\psi)
& = &
\big(\Mlt[K](c) \push \multinomial[K](\omega)\big)(\psi)
\\
& = &
\Mlt[K](c) \push \multinomial[K](\omega) \models \indic{\psi}
\\
& = &
\multinomial[K](\omega) \models \Mlt[K](c) \pull  \indic{\psi}.
\end{array} \]

\noindent Lemma~\ref{ValidityLem}~\eqref{ValidityLemInc} tells us that
in order to increase the latter validity we have to form the updated
distribution $\multinomial[K](\omega)\big|_{\Mlt[K](c) \pull
  \indic{\psi}}$, that we abbreviated as $\sigma$
in~\eqref{MultinomialUpdateAbr}. The next two results show that this
$\sigma$ is `close' to Jeffrey's update.

\begin{theorem}
\label{JeffreyFlrnThm}
Let $c\colon X \chanto Y$ be a channel with distribution $\omega\in\Dst(X)$
and data $\psi\in\Mlt(Y)$. Then:
\[ \begin{array}{rcl}
\flrn \push \Big(\multinomial[K](\omega)\big|_{\Mlt[K](c) \pull \indic{\psi}}\Big)
& = &
c_{\omega}^{\dag} \push \flrn(\psi).
\end{array} \]
\end{theorem}

\begin{myproof}
By the following argument.
\[ \begin{array}[b]{rcll}
\flrn \push \Big(\multinomial[K](\omega)\big|_{\Mlt[K](c) \pull \indic{\psi}}\Big)
& \smash{\stackrel{\eqref{DaggerEqn}}{=}} &
\Big(\flrn \klafter \Mlt[K](c)_{\multinomial[K](\omega)}^{\dag}\Big)(\psi)
\\[+0.5em]
& = &
\Big(\flrn \klafter \acc \klafter \big(c_{\omega}^{\dag}\big)^{K} 
   \klafter \arr\Big)(\psi)
   \qquad & \mbox{by Proposition~\ref{DaggerOfExtensionProp}}
\\[+0.5em]
& = &
\Big(\flrn \klafter \Mlt[K]\big(c_{\omega}^{\dag}\big) \klafter \acc
   \klafter \arr\Big)(\psi) \quad
   & \mbox{by Lemma~\ref{ExtensionLem}~\eqref{ExtensionLemNat}}
\\[+0.5em]
& = &
\big(c_{\omega}^{\dag} \klafter \flrn\big)(\psi)
   & \mbox{again by Lemma~\ref{ExtensionLem}~\eqref{ExtensionLemNat}}
\\[+0.5em]
& = &
c_{\omega}^{\dag} \push \flrn(\psi).
\end{array} \eqno{\QEDbox} \]
\end{myproof}

\subsection{Jeffrey's Rule as Variational Inference}\label{JeffreyArgminSubsec}

Variational inference~\cite{mackay2003information} is a well-known
technique in probability theory for finding approximations $\tau$ of
`difficult' distributions $\sigma$.  One then determines another
distribution $\tau$ as the distribution (from a certain class) that
diverges minimally from $\sigma$.

\begin{lemma}
\label{VILem}
Let an arbitrary distribution $\sigma\in\Dst\big(\Mlt[K](X)\big)$ be
given. The distribution $\omega\in\Dst(X)$ with minimal
Kullback-Leibler divergence
\[ \DKL\big(\sigma, \, \multinomial[K](\omega)\big) \]

\noindent is $\flrn \push \sigma \in \Dst(X)$.
\end{lemma}

\begin{myproof}
We first note that $\flrn \push \sigma \in \Dst(X)$ is given by:
\[ \begin{array}{rcccl}
\big(\flrn \push \sigma\big)(x)
& = &
\displaystyle\sum_{\varphi\in\Mlt[K](X)} \, \sigma(\varphi) \cdot \flrn(\varphi)(x)
& = &
\displaystyle\frac{1}{K} \cdot \sum_{\varphi\in\Mlt[K](X)} \, 
   \sigma(\varphi) \cdot \varphi(x).
\end{array} \eqno{(*)} \]

\noindent Then, for an arbitrary $\omega\in\Dst(X)$, we unravel the
divergence in the following manner, where $\mathsl{Const}$ is an
irrelevant constant that depends only on $\sigma$, not on $\omega$.
\[ \begin{array}{rcl}
\DKL\big(\sigma, \, \multinomial[K](\omega)\big)
& \smash{\stackrel{\eqref{KLEqn}}{=}} &
\displaystyle\sum_{\varphi\in\Mlt[K](X)} \, \sigma(\varphi) \cdot
   \ln\left(\frac{\sigma(\varphi)}{\multinomial[K](\omega)(\varphi)}\right)
\\[+1.2em]
& \smash{\stackrel{\eqref{MultinomialEqn}}{=}} &
\displaystyle\sum_{\varphi\in\Mlt[K](X)} \, 
   \sigma(\varphi) \cdot \ln\Big(\sigma(\varphi)\Big) -
   \sigma(\varphi) \cdot \ln\Big(\coefm{\varphi}\Big) -
   \sigma(\varphi) \cdot \sum_{x\in X} \varphi(x)\cdot\ln\Big(\omega(x)\Big)
\\[+1.2em]
& = &
\displaystyle \mathsl{Const} -
   \sum_{x\in X} \left(\sum_{\varphi\in\Mlt[K](X)} \, 
   \sigma(\varphi) \cdot \varphi(x)\right)\cdot\ln\Big(\omega(x)\Big)
\\[+1.2em]
& \smash{\stackrel{(*)}{=}} &
\displaystyle \mathsl{Const} - 
   K\cdot \sum_{x\in X} \big(\flrn \push \sigma\big)(x) \cdot
   \ln\Big(\omega(x)\Big)
\\[+1.2em]
& = &
\displaystyle \mathsl{Const} - 
   K\cdot \ln\left(\,\prod_{x\in X} \omega(x)^{(\flrn \push \sigma)(x)}\right).

\end{array} \]

\noindent Thus, in order to minimise the original divergence
$\DKL\big(\sigma, \multinomial[K](\omega)\big)$ we have to maximise
the latter log-expression $\ln\big(\cdots\big)$. This is a familiar
maximal likelihood estimation (MLE) problem, see
\textit{e.g.}~\cite[Ex.~17.5]{KollerF09}. The log expression is
maximal for $\omega = \flrn \push \sigma$. \QED
\end{myproof}

With this lemma we can get our `variational' characterisation of
Jeffrey's theorem.

\begin{theorem}
\label{JeffreyArgminThm}
Consider a channel $c\colon X \chanto Y$ with distribution
$\omega\in\Dst(X)$ and data $\psi\in\Mlt(Y)$. Jeffrey's update
$c_{\omega}^{\dag} \push \flrn(\psi)$ is the distribution
$\omega'\in\Dst(X)$ such that $\multinomial[K](\omega')$ diverges
minimally from multinomial update
$\multinomial[K](\omega)\big|_{\Mlt[K](c) \pull \indic{\psi}}$, that
is:
\[ \begin{array}{rcl}
\argmin\limits_{\omega'\in\Dst(X)} 
   \DKL\Big(\multinomial[K](\omega)\big|_{\Mlt[K](c) \pull \indic{\psi}}, \,
   \multinomial[K](\omega')\Big)
& = &
c_{\omega}^{\dag} \push \flrn(\psi).
\end{array} \]
\end{theorem}

\begin{myproof}
By Lemma~\ref{VILem} this minimal distribution is
\[ \flrn \push \Big(\multinomial[K](\omega)\big|_{\Mlt[K](c) \pull \indic{\psi}}\Big). \]

\noindent By Theorem~\ref{JeffreyFlrnThm} this equals Jeffrey's update
$c_{\omega}^{\dag} \push \flrn(\psi)$. \QED
\end{myproof}

\section{Conclusions}\label{ConclusionSec}

The difference in outcomes of Pearl's and Jeffrey's update rules
remains an intriguing topic. The paper does not offer the definitive
story about when to use which rule, but it does enrich the field with
several new ingredients (such as the different likelihoods and
variational inference) and offers a wider perspective (including
probabilistic programming). The main points that we have made explicit
are that, when we learn from data,
\begin{itemize}
\item repeated application of Pearl's rule, for each data point,
  corresponds to an update of the prior distribution, along a
  multinomial channel, see Theorem~\ref{PearlLikelihoodThm};

\item Jeffrey's rule is best understood as an update of all the
  multinomial draws from the prior and the formulation in Jeffrey's
  rule is a best approximation of this update, see
  Theorem~\ref{JeffreyArgminThm}.
\end{itemize}

\noindent In these two update mechanisms there seems to be different
perspectives at stake: the Pearlian posterior disease probability for
an \emph{individual} can be computed from a couple of tests, whereas
the Jeffreyan posterior probability for a \emph{population} requires
many tests.


\section{Appendix: Overview of WebPPL}\label{sec:webppl_intro}

\lstset{style=webppl}
We give a brief overview of the WebPPL probabilistic programming language: WebPPL is based on a purely functional subset of Javascript, that is then extended with probabilistic primitives for sampling, conditioning and inferring posterior probability distributions. Its implementation is described in detail \cite{dippl}, and the language can be tried out a browser under \url{webppl.org}. \\

\noindent In WebPPL, probability can be manipulated in the form of distribution objects (such as \lstinline|Bernoulli({p: 0.3})|) and as samplers, that is functions which return random draws from a distribution, such as \lstinline|bernoulli({p: 0.3})|. A distribution object \lstinline|dist| can be sampled from using the command \lstinline|sample(dist)|. Thus, the programs \lstinline|bernoulli({p: 0.3})| and \lstinline|sample(Bernoulli({p: 0.3}))| are equivalent. A shorthand for a biased coin flip is \lstinline|flip(p)|. WebPPL comes with a library of common probability distributions, both discrete and continuous. \\

\noindent The command \lstinline|condition(p)| expresses a boolean condition which must be met. The precise semantics of \lstinline|sample| and \lstinline|condition| will depend on the chosen inference algorithm, described below. Soft conditioning is available using the syntax \lstinline|observe(dist, observation)| but we don't need this in the current paper. \\

\noindent The command \lstinline|Infer(fn)| takes a sampler \lstinline|fn|, i.e. a higher-order function which represents a probabilistic experiment including conditions, and turns it into a distribution object which represents the exact or approximate posterior. The inference algorithm can be customized using a \lstinline|method| argument. The default algorithm for discrete problems such those as in this paper is exact enumeration. That is \lstinline|Infer| exhaustively tracks all random calls made within \lstinline|fn|, discards those that violate the conditions, and computes the exact posterior. This strategy is only feasible for small problem instances. A typical call to \lstinline|Infer| looks like
\begin{lstlisting}
var posterior = Infer({method: 'enumerate'}, function() {
  var x = bernoulli({p: 0.3})
  var y = bernoulli({p: 0.9})
  condition(x==y)
  return x
})
\end{lstlisting}
The result is a distribution object \lstinline|posterior|, which we can for example visualize using the command \lstinline|viz(posterior)|. We can also sample from the posterior using \lstinline|sample(posterior)|. Because \lstinline|Infer| and \lstinline|sample| are first-class operations in WebPPL, inference code can be nested without issue, expressing inference about inference. We use this pattern in our explanation of Jeffrey's update. \\

\noindent If the inference problems are no longer tractable using exact enumeration, approximate or sampling-based inference techniques can be used. The simplest is Monte Carlo simulation using rejection sampling, which will simply generate many execution traces of \lstinline|fn()|, discard those whose conditions haven't been satisfied, and aggregate the results. More sophisticated algorithms are importance sampling, particle filters, variational inference and Markov Chain Monte Carlo. Internally, WebPPL is compiled into continuation-passing style which allows the \lstinline|Infer| method a large amount of control over what happens at individual \lstinline|sample| and \lstinline|condition| commands \cite{dippl}.

\end{document}

%% file: causality.tikzstyles
\tikzstyle{white dot}=[inner sep=0mm, minimum size=1.5mm, draw=black, shape=circle, text depth=-0.2mm, draw=black, fill=white, tikzit category=nodes]
\tikzstyle{black dot}=[inner sep=0mm, minimum size=1.5mm, draw=black, shape=circle, draw=black, fill=black, tikzit category=nodes]
\tikzstyle{observed}=[inner sep=0mm, minimum size=5mm, draw=black, shape=circle, text depth=-0.2mm, draw=white, tikzit draw=gray, fill=white, tikzit category=dag]
\tikzstyle{latent}=[inner sep=0mm, minimum size=5mm, draw=black, shape=circle, text depth=-0.2mm, draw=black, fill=white, tikzit category=dag]
\tikzstyle{small box}=[shape=rectangle, text height=1.5ex, text depth=0.25ex, yshift=0.5mm, fill=white, draw=black, minimum height=6mm, yshift=-0.5mm, minimum width=6mm, font={\small}, tikzit category=boxes]
\tikzstyle{medium box}=[shape=rectangle, draw=black, fill=white, small box, minimum width=8mm, tikzit category=boxes]
\tikzstyle{semilarge box}=[shape=rectangle, draw=black, fill=white, small box, minimum width=12.5mm, tikzit category=boxes]
\tikzstyle{large box}=[shape=rectangle, draw=black, fill=white, small box, minimum width=15mm, tikzit category=boxes]
\tikzstyle{upground}=[circuit ee IEC, thick, ground, rotate=90, scale=1.5, inner sep=-2mm, tikzit shape=circle, tikzit fill=blue, tikzit category=points]
\tikzstyle{downground}=[circuit ee IEC, thick, ground, rotate=-90, scale=1.5, inner sep=-2mm, tikzit shape=circle, tikzit fill=green, tikzit category=points]
\tikzstyle{point}=[regular polygon, regular polygon sides=3, draw, scale=0.75, inner sep=-0.5pt, minimum width=9mm, fill=white, regular polygon rotate=180, tikzit category=points]
\tikzstyle{copoint}=[regular polygon, regular polygon sides=3, draw, scale=0.75, inner sep=-0.5pt, minimum width=9mm, fill=white, tikzit category=points]
\tikzstyle{uniform}=[point, fill=gray, tikzit shape=circle, scale=0.5]
\tikzstyle{label}=[font={\footnotesize}, text height=1.5ex, text depth=0.25ex, tikzit draw=blue, tikzit fill=white, tikzit category=labels]
\tikzstyle{left label}=[label, anchor=east, xshift=2mm, tikzit draw=green, tikzit fill=white, tikzit category=labels]
\tikzstyle{right label}=[label, anchor=west, xshift=-2mm, tikzit draw=purple, tikzit fill=white, tikzit category=labels]
\tikzstyle{disintegration}=[draw=black, fill={gray!50}, tikzit fill=gray, shape=rectangle, minimum width=1.6cm, minimum height=1.2cm, opacity=0.3]
\tikzstyle{empty diag}=[shape=rectangle, draw=darkgray, dashed, minimum width=8mm, minimum height=8mm, yshift=0.5mm]

\tikzstyle{diredge}=[->, >=latex]
\tikzstyle{dashed edge}=[-, dashed]
\tikzstyle{plate}=[-, draw=black, thick]

%% file: jeffrey-pearl.tikz
\begin{tikzpicture}[scale=0.4]
	\begin{pgfonlayer}{nodelayer}
		\node [style=point] (0) at (0.75, -2) {$\omega$};
		\node [style=small box] (1) at (0.75, 1) {$c$};
		\node [style=none] (2) at (0.75, 7) {};
		\node [style=none] (3) at (2.75, -3) {};
		\node [style=none] (4) at (-1.25, -3) {};
		\node [style=none] (5) at (-1.25, 5) {};
		\node [style=none] (6) at (2.75, 5) {};
		\node [style=none] (7) at (0.75, 8) {$\Mlt(Y)$};
		\node [style=none] (10) at (9.5, 7) {};
		\node [style=none] (15) at (9.5, 8) {$\Mlt(Y)$};
		\node [style=point] (16) at (7.5, -2) {$\omega$};
		\node [style=small box] (17) at (7.5, 1) {$c$};
		\node [style=point] (18) at (11.5, -2) {$\omega$};
		\node [style=small box] (19) at (11.5, 1) {$c$};
		\node [style=small box] (20) at (9.5, 4.25) {$\quad\acc\quad$};
		\node [style=none] (21) at (8.5, 4) {};
		\node [style=none] (23) at (10.5, 4) {};
		\node [style=none] (24) at (5, 1) {=};
		\node [style=none] (25) at (-0.25, -0.75) {$X$};
		\node [style=none] (26) at (-0.25, 3) {$Y$};
		\node [style=none] (27) at (9.5, 1) {$\cdots$};
		\node [style=point] (28) at (24.75, -2.5) {$\omega$};
		\node [style=small box] (29) at (24.75, 3) {$c$};
		\node [style=none] (30) at (24.75, 8.5) {};
		\node [style=none] (31) at (27.75, 0) {};
		\node [style=none] (32) at (21.75, 0) {};
		\node [style=none] (33) at (21.75, 6) {};
		\node [style=none] (34) at (27.75, 6) {};
		\node [style=none] (35) at (23, 8.5) {$\Mlt(Y)$};
		\node [style=none] (36) at (23.75, 1) {$X$};
		\node [style=none] (37) at (23.75, 5) {$Y$};
		\node [style=point] (38) at (34.25, -2) {$\omega$};
		\node [style=none] (40) at (34.25, 8.5) {};
		\node [style=small box] (41) at (32.25, 3) {$c$};
		\node [style=small box] (42) at (36.25, 3) {$c$};
		\node [style=none] (43) at (34.25, 3) {$\cdots$};
		\node [style=black dot] (45) at (34.25, 0.5) {};
		\node [style=none] (46) at (33.25, 6) {};
		\node [style=none] (48) at (35.25, 6) {};
		\node [style=medium box] (49) at (34.25, 6.5) {$\quad\acc\quad$};
		\node [style=none] (50) at (29.75, 3) {=};
		\node [style=none] (51) at (33.25, -0.5) {$X$};
		\node [style=none] (52) at (32.5, 8.5) {$\Mlt(Y)$};
		\node [style=none] (53) at (6.5, -0.75) {$X$};
		\node [style=none] (54) at (10.5, -0.75) {$X$};
	\end{pgfonlayer}
	\begin{pgfonlayer}{edgelayer}
		\draw (0) to (1);
		\draw (1) to (2.center);
		\draw [style=plate] (3.center) to (4.center);
		\draw [style=plate] (4.center) to (5.center);
		\draw [style=plate] (5.center) to (6.center);
		\draw [style=plate] (6.center) to (3.center);
		\draw (16) to (17);
		\draw (18) to (19);
		\draw [in=-90, out=90, looseness=1.50] (17) to (21.center);
		\draw [in=-90, out=90, looseness=1.50] (19) to (23.center);
		\draw (20) to (10.center);
		\draw (28) to (29);
		\draw (29) to (30.center);
		\draw [style=plate] (31.center) to (32.center);
		\draw [style=plate] (32.center) to (33.center);
		\draw [style=plate] (33.center) to (34.center);
		\draw [style=plate] (34.center) to (31.center);
		\draw (45) to (38);
		\draw [in=-90, out=150, looseness=1.25] (45) to (41);
		\draw [in=-90, out=30, looseness=1.25] (45) to (42);
		\draw [in=-90, out=90, looseness=1.50] (41) to (46.center);
		\draw [in=-90, out=90, looseness=1.50] (42) to (48.center);
		\draw (40.center) to (49);
	\end{pgfonlayer}
\end{tikzpicture}